
\magnification=1200
\baselineskip=14pt
\font\gran=cmr10 scaled \magstep 2
\font\sc=cmcsc10 scaled \magstep 1
\line{\hfil UB-ECM-PF 92/1 }
\line{\hfil January 1992 }
\vskip 3.2truecm
\centerline{\gran  Spectrum of Relativistic Fermions in a 2d Doped
Lattice}
\vskip 1.2truecm
\centerline{{\sc D. Espriu} and
{\sc J. Matias}}
\medskip
{\it
\centerline{ D.E.C.M.}
\centerline{ Universitat de Barcelona }
\centerline{ Diagonal, 647}
\centerline{ E-08028 Barcelona}}
\vskip 1.8truecm
\centerline{      ABSTRACT  }
\medskip
Motivated by some previous work on fermions on random lattices and
by suggestions that impurities could trigger parity breaking in 2d
crystals, we have analyzed the spectrum of the Dirac equation on
a two dimensional square
lattice where sites have been removed randomly
--- a doped lattice. We have found that the system is well described
by a sine-Gordon action. The solitons of this model are the
lattice fermions, which pick a quartic interaction due to the doping
and become Thirring fermions. They also get an effective mass different
from the lagrangian mass. The system seems
to exhibit spontaneous symmetry
breaking, exactly as it happens for a randomly triangulated lattice.
The associated ``Goldstone boson" is the sine-Gordon scalar. We argue,
however, that the peculiar
behaviour of $\langle \bar{\psi}\psi\rangle$ is due to finite size
effects.

\vfill
\eject


Sometime ago the spectrum of the Dirac equation on a
two dimensional random lattice was studied in great
detail[1]. It was then found numerically that the density of
eigenvalues $\rho(\lambda)$
on randomly triangulated lattices exhibits
a rather peculiar behaviour. Instead of behaving linearly
for small values of the eigenvalue $\lambda$,
$\rho(\lambda) \sim \lambda$, as it does in the
continuum and also in a regular lattice (albeit with an
additional factor of 4 due to doubling), the density of
eigenvalues does not vanish as $\lambda\to 0$
for the lattices considered.
Indeed it would seem that, as the lattice size increased,
$\rho(\lambda)$ would go to a constant value or even
diverge as $\lambda\to 0$.
Accordingly, the fermion
condensate, which on general grounds is related to the
spectral density
$$\langle \bar{\psi}\psi\rangle =\pi \rho(0)\eqno(1)$$
would not vanish. Since we are in a two dimensional
system which is not supposed to undergo spontaneous
symmetry breaking[2],
 the behaviour of $\langle \bar{\psi}\psi\rangle$
in a random lattice
is quite unexpected.
Of course from the numerical
evidence alone
one cannot {\it rigorously} conclude that
$$
 \lim_{m\rightarrow 0}\lim_{V\rightarrow\infty}
\langle\bar\psi\psi\rangle
 \neq 0 \eqno(2)
$$
since one is forced to work with finite lattices, but
the results of [1] strongly suggest that in
the limit (2)  $\langle\bar\psi\psi\rangle$ tends to a finite constant.
This non-zero value for $\langle\bar\psi\psi\rangle$ is clearly due
to the ``interactions" induced by the randomness on the otherwise
free fermions.
\medskip

It was also observed in [1] that the pseudoscalar Green
function $\langle (\bar\psi\gamma^5\psi)(x)
(\bar\psi\gamma^5\psi)(0)\rangle$
has, in addition to a pole at twice the mass of the fermion,
an unmistakeable signal of another
particle in the spectrum with a mass $m_{55}\sim \sqrt{m/a}$.
The fermion mass itself is not affected by the randomness in
any noticeable way. The extra particle seen in the pseudoscalar
channel
would correspond to the (pseudo) ``Goldstone boson"  of the broken
$U(1)_A$ symmetry. The reason for the quotation marks is that
no genuine Goldstone boson is
supposed to exist in 2d due to the infrared problems characteristic
of 2d physics.
 However, we never take $m\to 0$ in the
random lattice because we always work with finite volumes. In those
conditions the particle found in [1] behaves as a pseudo Goldstone
boson, so the question of whether, in the appropriate limits,
the Goldstone
goes away or not is somewhat academic. The fact is that physics on
a 2d finite random lattice
is very well accounted for by assuming that
{\it there is} spontaneous symmetry breaking, with all its
consequences.
The particle found in [1]
 is, however, unstable; it decays into
two fermions.
\medskip

The other remarkable feature of the random lattice is that
the doubled fermions go away, at least for the free theory[1,3].
 Indeed, the
correlators on the random lattice behave at
long distances exactly as their continuum counterparts, normalization
included. The doubles have acquired a very large dynamically generated
mass and have disappeared from the spectrum. The only trace of the
would be doubles is the ``Goldstone boson"  discussed in the
previous paragraph.
The anomaly is
also well reproduced. This is possible because the random
lattice bypasses one of the hypothesis of the Nielsen-Ninomiya
theorem[4]. In practice,
for interacting theories some mass fine tuning turns out to be
necessary in a way somewhat similar to what has to be done for
Wilson fermions[3].
One's hope is that some of the
features observed in [1] (in particular the appearance of
$\langle\bar\psi\psi\rangle
 \neq 0 $)
must be, in some sense, universal.
Relativistic fermions on a regular two dimensional doped lattice is
a natural, more tractable, alternative.

\medskip

There is also another reason to study relativistic fermions
in a doped lattice. Many interesting materials in condensed
matter physics (in particular those where
high $T_c$ superconductivity  has been described)  are two dimensional
in nature. One could well think of a regime in which some
quenched impurities prevent electrons (or holes) from
occupying some lattice sites.
If we mimic this potential by some
repulsive potential the corresponding Schr\"odinger
equation would be
$$i{\partial\over \partial t}\phi=-{1\over 2m}\Delta^2 \phi
+ V \phi\eqno(3)$$
where $\Delta$ is the lattice laplacian and $V $ vanishes if the
site is not occupied and is large and positive if there's
an impurity in the site in question. We will not
consider the relevant case of having a finite electron density, neither
interactions amongst them.
It turns out that the physical situation described by (3) can be
obtained as the non-relativistic limit of the Dirac
equation
$$ i{\partial \over \partial t}\psi(x,t) =
-i\alpha_i \Delta_i \psi(x,t) + m \sigma_3 \omega(x) \psi(x,t)
\eqno(4)$$
provided that we let $w(x)\to \infty$ at the impurity sites.
$\Delta_i$ is the lattice derivative in the $i$ direction,
while the $t$ direction is kept continuous.
The eigenstates and eigenvalues
of the 2 dimensional Dirac equation correspond
to the energy eigenvalues and eigenstates of the 3d
quantum hamiltonian.
This  problem is exactly equivalent to
a lattice where the impurity sites have
been removed. In order to see this we simply have to
scale the fields to absorb
$w(x)$. The resulting lattice Dirac operator will be
$$D(x,y)={\gamma_\mu l_\mu (x,y) U(x,y)
\over 2\sqrt{w(x)w(y)} } + m \delta(x,y)\eqno(5)$$
where $l_\mu (x,y)$ are the lattice vectors and
$U(x,y)$ is a gauge field that we have added for
later use.
The only difference
between the spectrum obtained by finding the eigenvalues of
(5) and then going to the non-relativistic limit
 and the Schr\"odinger problem described
by (3) would be that
 squaring $D$ produces a laplacian
connecting sites at distance two. This is of course well known
and it just tells us that we will have four species.
In contrast to a random lattice, doping a regular lattice
is not enough
to remove the doubling.
\medskip

In what follows
we will see that free fermions in a two dimensional doped
lattice  exhibit, even
for small masses and large volumes, a large value for
$\langle \bar\psi\psi\rangle$, a phenomenon
closely resembling the one taking place in a random lattice.
We will also see that there is a pseudo Goldstone boson too.
As we pointed out before,
both observations are clearly at odds with the common lore.
However, the numerical results look
very convincing: a person
who has never heard about
Mermin-Wagner's  or Coleman's theorem would conclude that indeed
there is spontaneous symmetry breaking in this 2d system.
Yet, we have reasons to believe that the appearance of
spontaneous symmetry breaking in the doped lattice is, rather
paradoxically, a finite size effect.
As we will show, the
complete system, fermions plus Goldstone
boson, is well described for any value of the mass
by a massive Thirring model[5] and
its bosonic counterpart, the sine-Gordon model[6]. The Thirring
fermions will correspond to our original lattice fermions, the
doping having induced  a quartic chiral invariant interaction.
Using this information we will see that the symmetry breaking
disappears in the infinite volume limit. We will also see
that there are more differences between a random lattice
and a doped lattice than we originally thought.

\medskip

Our system consists of a two dimensional square regular lattice with
lattice spacing $a$ where we have taken out randomly
points of the space, to simulate impurities in a crystal.
The lattice is basically a regular square lattice where
pairs of points separated a distance $a$ are joined by a link if none
of the points of the couple is an impurity and otherwhise there is no
link. The construction of the dual lattice is straightforward except for
the dual cells associated to the points near an
impurity which have to be modified to make sure the sum of
areas is equal to the complete
volum of the system. There is some freedom in doing this which does
not seem to matter. The
symmetries of the problem ($C$, charge conjugation, and
chirality, $\gamma_{5}$) tell us that the eigenvalues come in sets
of four.
Parity and discrete rotations are not exact
symmetries any longer. They have been broken by the doping;
their restoration in an broader statistical sense has to be checked
explicitly (they are indeed restored).
 \medskip

 Let us begin by evaluating the
chiral condensate in our doped lattice.
The condensate is defined by
$$ \langle\bar\psi\psi\rangle={1\over V} \sum_{x=1}^{V} \omega(x)
 tr (D+m)^{-1}(x,x) =
{1\over V} \sum_{\lambda \geq 0} {2m
\over \lambda ^{2} + m^{2}}\eqno(6) $$
We have computed this expression using both periodic and
 antiperiodic boundary
conditions for the fermion fields without finding any
noticeable differences.
We have considered
 several lattice sizes
and different densities of impurities.
In all the cases we have examined $\langle \bar\psi\psi\rangle$
is much bigger than in the regular lattice case, showing
a behaviour which is not too different from the
random lattice. This is illustrated in Fig. 1. The
behaviour of $\langle \bar\psi\psi\rangle$ strongly
suggests that the $U(1)_A$ symmetry of free fermions is
spontaneously broken by the onset of randomness. The breaking
appears to be stronger for higher
density of impurities. For smallish sizes $\langle \bar\psi\psi\rangle$
fluctuates rather wildly from a given lattice to another, but
beyond
some size the curves order nicely. Notice that for a given lagrangian
mass $\langle\bar\psi\psi \rangle$ seems to decrease slowly
with the volume.

\medskip

Has the fermion mass been somehow modified by the doping?
The zero momentum fermion propagator will
give this information to us. It is defined by
$$
   G^{\alpha \beta}(x)={1 \over V} \int dx_{1}\, dy_{1} \, dx_{2}
\, dy_{2}\, \langle
\bar\psi^{\beta}(x_{1}, y_{1})  \psi^{\alpha}(x_{2},  y_{2})\rangle
\delta (x-(x_{1}-x_{2})) \eqno(7)
$$
where we have averaged over vertical bins of width $a/2$ or slightly
greater if there is an impurity on the left or right side of the bin
we are averaging. Using symmetry arguments we can
write $ G(x)=A(x)1+B(x)\gamma_1 $,
so $A(x)+B(x)$ describes a particle
propagating forward in time. In a
regular lattice of size $ V=L_{1} \times L_{2} $
$$
A(x)+B(x)={1 \over L_{1}} \sum_{n=1}^{L_{1}} { \sin p
\sin p x + Ma \cos p x \over (Ma)^{2}+ \sin^2 p}
  \qquad p={2\pi n\over  L_{1}}\eqno(8)
 $$
(We have mostly used periodic boundary conditions for these fits.)
Comparing this expression with the curve obtained numerically by
inverting the Dirac operator
one obtains  the mass of
the propagating particle. The fits are always extremely good,
except for the smaller masses (typically below $m=0.05$)
where performing a quenched average over a set of lattices
may be required.
We have found that, unlike in a
random lattice,
 the fermion mass observed
in the propagator, $ M$, is no longer the lagrangian mass $m$.
The propagator mass grows
linearly with the lagrangian mass for large masses, but with
a coefficient larger than one. This coefficient, in addition,
increases with the doping but seems to be volume independent.
For small masses, on the other hand,
 there is some saturation at an almost constant
value, as can be observed in Fig. 2, suggesting that in the
$m\to 0$ limit the propagator mass $M(m)$ stays finite. The finite
value to what $M$ tends as $m\to 0$, $M(0)$, clearly decreases with
increasing volumes, indicating that it may be due to
finite size effects. $M(0)$ grows with the
density of impurities.
In addition to a mass renormalization induced by the doping,
there is also some wave function renormalization.
The mass and wave function corrections
obtained from the fermion propagator coincide
very nicely
with the ones that come from the connected part of the scalar propagator
$P_{11}(x)=\langle(\bar\psi\psi)(x)(\bar\psi\psi)(0)\rangle$.
In the absence of other particles in this channel this is
expected to decay as $2M$ and it does. The agreement is really good,
as Fig. 3 shows.
{}From measuring this
Green function we also find out that the doubles subsist, the
overall normalization of $P_{11}$
being approximately four times the naive one.
In some sense our system resembles
a random lattice
but it is still ordered enough to allow  the survival of the
doubles.
\medskip

Is there a pseudo ``Goldstone boson" ?   Let's examine the
pseudoscalar propagator
$$
   P_{55}(x)={1 \over V} \int dx_{1}\, dy_{1}\, dx_{2}\, dy_{2}\,
\langle (\bar\psi \gamma^5 \psi)(x_{1}, y_{1})
(\bar\psi \gamma^5 \psi )(x_{2}, y_{2})\rangle
\delta (x-(x_{1}-x_{2}))\eqno(9)
 $$
To calculate this correlator we invert the Dirac operator,
using periodic boundary conditions,
for different masses and densities. By plotting this
zero momentum propagator we clearly see the contribution
from two type of particles. First of all, there is an
exponential decay (actually a cosh behaviour in our finite lattice)
with a mass $2M$, exactly as for the $P_{11}$ propagator.
This corresponds to the free propagation of the fermion -
antifermion pair in the lattice
and it is described by an expression of the form
$$
P_{55}^{reg}(x)=C {2 \over L_1^2 L_{2}} \sum_{n=1}^{L_1}
\sum_{m=1}^{L_1}
\sum_{k=1}^{L_2}
{((Ma)^{2} + \sin p_1 \sin p_2 + \sin^2 q)
 \cos (p_1-p_2)x
\over ((Ma)^2+ \sin^2 p_1 +\sin^2 q)
((Ma)^2+ \sin^2 p_2+ \sin^2 q)}
$$
where $p_1={2\pi n /  L_1} , p_2={2\pi m /
L_1} , q={ 2\pi k /  L_2} $ and $C$
is the additional wave function renormalization
factor we mentioned before. There are no free
parameters here, everything is obtained from the fermion
propagator.
This would be the only contribution in a regular lattice (but with
$M=m$).
 However,
once this behaviour has been subtracted from the measured
propagator there is still a clear signal of a  particle in the
pseudoscalar
channel. This signal
is perfectly adjusted by the propagator of a scalar
particle in a finite box,
with a coefficient that we fit. The fit to the
scalar particle, with
mass $m_{55}$,
plus
free fermions, with mass $M$,  really works perfectly (Fig. 3)
for all the desities of impurities we have tried (up to
20\%) and for the whole range of sizes.
\medskip

The above situation is again reminiscent of the random
lattice. However, here the scalar particle, that we
interpret as the pseudo ``Goldstone boson"  of the ``broken"
symmetry, is actually the lighter state in its channel and
therefore stable, $m_{55} < 2M$.
 Its mass $m_{55}$ as a function of the
lagrangian mass $m$ is plotted in Fig. 4. For large
values of $m$, $m_{55}$ grows linearly with $m$.
For smaller values there are deviations with
respect to this behaviour. We have found that $m_{55}$
is extremely well adjusted by the formula
$$m_{55}^2 = c_1 m^b + c_2 m^2 \qquad, \qquad b<1 \eqno (11) $$
In the random lattice, on the
contrary, the behaviour is $m_{55}^2\sim m$ to a very good
approximation.
Very small values of the
mass were not investigated in [1].
\medskip

Let us now try to gain some theoretical understanding of the
above results. We have seen that $\langle\bar\psi\psi\rangle$
is large in our system.
On general grounds we then expect that the generator
of the broken symmetry acting on the vacuum state produces
a Goldstone-like particle and indeed this particle is clearly
seen in the pseudoscalar channel. Should the symmetry be
exact, the ``Goldstone" would be exactly massless. However there is
a explicit breaking of the form
$$ m\bar\psi_R\psi_L + m\bar\psi_L\psi_R\eqno(12)$$
The bosonization rules in two dimensions
force us to make the identifications
$$ \bar\psi_L\psi_R\sim e^{-i{\sqrt{2}\over f}\phi}\quad\quad
\bar\psi_L\psi_R\sim e^{i{\sqrt{2}\over f}\phi}\eqno(13)$$
On general grounds we
expect for this particle a lagrangian
$${\cal L} = {1\over 2}({\partial_{\mu} \phi})^2 + {\hat
m}^2 { f^2 \over 2 }: \cos({\sqrt{2}\over f} \phi):
\eqno(14)$$
The sign of the second term is irrelevant; it can be absorbed
in a shift of $\phi$. The constants ${\hat m}$ and $f$
cannot be determined by symmetry arguments alone. We know, however,
that ${\hat m}$ must vanish when $m\to 0$. It is also easy
to see that for free fermions $f^2=1/2\pi$. Any departure from
this value will indicate that the fermions are not free. This will
indeed be the case for non-zero doping density.
\medskip

\medskip

Eq. 14 is of course nothing but the well known sine-Gordon
lagrangian. Sine-Gordon
is solvable at the classical level and its quantum spectrum can be
found using the WKB method[8] and from other considerations[9].
The exact spectrum consists of two type of
particles: solitons and antisolitons, with a mass $M_{S}=8 {\hat m}/
\gamma$,  and a finite number of
soliton-antisoliton bound states with masses
$ M_{n}=(16 {\hat m} / \gamma)\sin({n\gamma / 16}) $
where $ n=1,2,3,..<(8\pi / \gamma) $ and $ \gamma =
{2\over f^2}(1-{1\over 4\pi f^2})^{-1} $.
The solitons of this model
are the fermions of the Thirring model,  which as Coleman
demonstrated[7] long ago has the same spectrum as sine-Gordon,
and whose lagrangian is
$${\cal L}= \bar\psi\gamma_\mu\partial_\mu\psi+M_0\bar\psi\psi
-{g\over 2}\bar\psi\gamma_\mu\psi
\bar\psi\gamma_\mu\psi \eqno(15)$$
The physical mass of the fermions,
renormalized by the quartic interactions, is
$M_S$. The relation between the quartic Thirring coupling $g$
and $f$ is
${g/\pi}=2\pi f^2 - 1   $.
\medskip
One may worry to what extent the equivalence Thirring -
sine-Gordon holds on a lattice. Quite apart from general symmetry
arguments, let us recall that the equivalence is seen by
perturbing both theories around the massless case and comparing
Green functions[7]. It is easy to see that this equivalence holds
on a lattice too. The lattice massless propagators corresponding to
the operators $\bar\psi\psi$ in the l.h.s of (13) are, in the
infinite volume limit,
identical to the ones corresponding to their bosonic counterparts
in the r.h.s. of (13), but defined in a lattice with spacing $2 a$.
We
expect to have four of such bosons.
We have seen that the scalar particle in the $P_{55}$ channel
can be described by the sine-Gordon lagrangian. It is then natural
to identify the Thirring fermions
with our doped lattice fermions, whose
bound state in sine-Gordon (fermion-antifermion bound state in Thirring)
corresponds to our Goldstone boson.
The Thirring - sine-Gordon correspondence predicts the
mass ratio
$$ {m_{55}\over 2M} =\sin{n\gamma\over 16}\eqno(16)$$
Indeed, for all densities we always find this ratio
to be less than one. Note that since $\hat m$ contains
some normal ordering renormalization, $M_S$ and $M_n$ cannot
be unambiguosly determined, but the ratio is well defined.
We only observe one  single pole in the $P_{55}$ channel, so it
must be the case that  $\gamma > 4\pi$ (or, equivalently,
$f^ 2 < 3/4\pi$). When comparing the different spectra it should
be borne in mind that finite size effects affect Thirring and
sine-Gordon differently. These effects are expected to be smallest
for relatively large masses. Indeed we observe that the
ratio (16) is very approximately constant beyond some
mass and from this ratio we can extract the parameter
$f$ (or $g$) that characterizes the model.
For the doping densities 20, 10, 4.9\% we get $g=$1.06, 0.90, 0.83,
respectively.
Not surprisingly, the value of $g$ increases with the
doping concentration. In the
$m\to 0$ limit, the mass
of the Thirring fermion is substantially
modified by  finite size effects[10].
In the zero mass limit one expects $M\sim g^2/L$. This is
in good agreement with the ``saturation"
behaviour observed in Fig. 2.
The fermion-antifermion composite boson, on the
other hand, becomes free in that limit so we expect
$m_{55}\to 0 $ as $m\to 0$ even in a finite volume.
\medskip

Contrary to the doped case,
the random lattice has $m_{55}> 2M$. This implies
that the random lattice falls in a domain of parameters
where the correspondence sine-Gordon - Thirring fails.
(The effective potential of sine-Gordon is unbounded from
below for $f^2 <1/4\pi$.) Indeed,
in the random lattice the ``Goldstone"
decays into a pair of physical fermions with mass $m$.
Since in a random lattice
the correspondence with Thirring fails, we have no reason
to expect a four fermion term.
This was confirmed
in [1]; the physical fermions
are indeed free on a random lattice.
The doubles, however, have acquired
a mass of ${\cal O}(1/a)$ due to the randomness.
Another difference we have between the  random lattice case
and our doped lattice is the behaviour of $m_{55}$ as a function
of the mass. While we find a behaviour of the form (11), in the
random lattice a dependence of the form $m_{55}\sim \sqrt{m}$
describes well the data. All this suggests that
the randomly triangulated lattice exhibits, for a broad range of
masses,  a behaviour
much more similar to the one of a system undergoing
genuine spontaneous symmetry breaking.
\medskip

Now we would like to turn to $\langle\bar\psi\psi\rangle$
again. We have arrived at the conclusion that our doped
lattice is well described by Thirring / sine-Gordon. Since
the partition function of the latter can be computed in the
WKB approximation, by making a derivative with respect to $m$
we can estimate theoretically the contribution from
the bosonic sector of the theory to $\langle\bar\psi\psi\rangle$.
Adding the contribution from the solitonic sector ---just
a loop of fermions with mass $M$---
we can compute $\langle\bar\psi\psi\rangle$. We
will restrict ourselves to small
masses, where the use of a  continuum formalism for a scalar
is justified.
For masses $m$ comparable to the lattice spacing $a$ one would have
to estimate the partition function using consistently the lattice
action.
After evaluating the partition function
by the saddle point method
and taking care of the quadratic fluctuations around a doublet
state [8,11], one arrives at the following result for the
partition function in a box of
size $L\times T$ in Minkowski space
$$
{\rm Tr} e^{iHT}\simeq LT \sum_n \sum_r\sqrt{ M_r\over
(T^2-(nL)^2)^{3/2}} \exp(-iM_r\sqrt{T^2-(nL)^2})\eqno(17)$$
The sum over $r$ runs over all doublet states. As we have seen
we have only one. The above expression is remarkably simple.
Apart from some prefactors this is the first quantized action
for a relativistic scalar particle in a one-dimensional box
of size $L$, if we take into account that $nL/T$ is the velocity
$v$ of the particle (which may wrap around several times). $v$
(or $n$ ) is in fact a collective coordinate.
In euclidean space time we have to Wick rotate the coordinates
appropriately. We also have to include solutions that wrap in
the $T$ direction, to restore the $L\leftrightarrow T$
symmetry. So, finally
$$
Z_{SG}={\rm Tr} e^{-HT}\simeq LT \sum_n \sum_m\sqrt{ m_{55}\over
((mT)^2+(nL)^2)^{3/2}} \exp(-m_{55}\sqrt{(mT)^2+(nL)^2})\eqno(18)$$
The value of $\langle\bar\psi\psi\rangle$ is
obtained as %
$$\langle\bar\psi\psi\rangle=-{1\over LT}{ \partial\over
\partial m}\log Z\eqno(19)$$
For the sizes and masses we have been considering, the
leading contribution to the chiral condensate is, for a square lattice,
$$\langle\bar\psi\psi\rangle\simeq 4\times {1\over 2L}{\partial m_{55}
\over \partial m}\eqno(20)$$
where we have included the factor of 4 due to doubling. For small
masses $m_{55}\sim m^{b/2}$, with $b<1$,
so the chiral condensate behaves as
$1/Lm^{(1-b/2)}$. This expression clearly shows that the appearance
of $\langle\bar\psi\psi\rangle\neq 0$ in a doped lattice is a finite
size effect, at least for the contribution that comes from
the ``Goldstone boson". What about the contribution from the
solitonic sector?  This is proportional to $M\log M$, and since
$M$ vanishes in the infinite volume limit, it will vanish too.
Fig.1c shows the comparison between the values of
$\langle\bar\psi\psi\rangle$ obtained by numerical evaluation
in the doped lattice and analytically from Thirring / sine-Gordon in
the WKB approximation. The agreement is qualitatively good.
\medskip
To conclude we would like to mention several other points
related to our system.
\medskip

i) Since our doped
lattice  has doubled fermions one expects the chiral
anomaly to vanish. That this is the case it can be seen
by switching on the gauge field $U(x,y)$ and making
a perturbative expansion. The first order contribution
would require the evaluation of the Green function
$\langle (\bar\psi\gamma^5\psi)(0) (\bar\psi\gamma_\mu
\gamma^5\psi)(x)\rangle$, which is precisely the one
leading to the anomaly in two dimensions. It can be
seen easily that this quantity is identically zero.
\medskip

ii) A natural question to ask is whether the presence of
randomness has induced localization of some sort. The answer
is no. We have checked that all the eigenvectors of the
Dirac equation are extended states,  as far as one can tell
numerically. This is also at variance with the situation
in a randomly triangulated lattice where it was seen that some
of the states were localized.
\medskip

iii) It has been argued sometimes in the literature that
the net effect of impurities could be an additional
 breaking of parity (and
$T$ invariance) in the 2+1 quantum system, triggering the
emergence of a Chern-Simmons term in the effective action
with a non standard coefficient.
(The mass term itself breaks parity in 2+1 dimensions.) We
believe that the net effect of the impurities is simply
to renormalize the fermion mass.
\bigskip

This work has been supported by the CICYT grant AEN89-0347 and
the EEC Science Twinning Grant SCI-000337. J.M. acknowledges
the financial support of a FPI graduate fellowship from the Ministerio
de Educaci\'on y Ciencia.

\bigskip
\line{\bf References\hfil }
\medskip
\item{[1]}{D.Espriu, M.Gross, P.Rakow and J.F.Wheater,
           Nucl. Phys. B275 (1986) 39;
D.Espriu, M.Gross, P.Rakow and J.F.Wheater,
        Prog. Theor. Phys. (Supp.) 86 (1986) 304}
\item{}{
See also: R.Friedberg, T.D.Lee and H.Ren,
Prog. Theor. Phys. (Supp.) 86 (1986) 322;
Y.Pang and H.Ren, Phys. Lett. B172 (1986); 195 (1987) 223;
H.Ren, Nucl. Phys. B300 (1988) 531}
\item{[2]}{S.Coleman, Comm. Math. Phys. 31 (1973) 259  }
\item{[3]}{S.Perantonis and J.F.Wheater, Nucl. Phys.
       B295 (1988)443    }
\item{[4]}{H.B.Nielsen and N.Ninomiya, Nucl. Phys. B185 (1981)20;
B193 (1981) 173}
\item{[5]}{W.Thirring, Ann. Phys. (N.Y.), 3 (1958) 91}
\item{[6]}{See e.g. R. Rajaraman, Phys. Rep. C5 (1975) 227,
      and references therein   }
\item{[7]}{S.Coleman, Phys. Rev. D11 (1975) 2088}
\item{[8]}{R.F.Dashen, B.Hasslacher and A.Neveu, Phys. Rev.
D11 (1975) 3424}
\item{[9]}{A.B.Zamolodchikov and Al.B.Zamolodchikov,
Ann. Phys. (N.Y.) 120
(1979) 253 }
\item{[10]}{H.Yokota, Prog. Theor. Phys. 77 (1987)
1450}
\item{}{F.Ruiz and R.F.Alvarez-Estrada, Phys. Lett. B 182 (1986)
   354}
\item{[11]}{R.F.Dashen, B.Hasslacher and A.Neveu, Phys. Rev.
D10 (1974) 4114}

\vfill
\eject
\line{\bf Figure Captions\hfil}
\medskip
\item{Fig. 1.-} (a.) {$\langle\bar\psi\psi\rangle$
 as a function of the
lagrangian mass for three volumes: (${V_1}$) 648, (${V_2}$) 1656 and
(${V_3}$) 2304. The density of impurities is 10\% . The dashed line is
the value of $\langle\bar\psi\psi\rangle$ in a regular lattice.
The value in the doped lattice is much bigger, particularly for
small masses, suggesting that it is sensible to use the term
``spontaneous symmetry breaking". The value of $\langle
\bar\psi\psi\rangle$ is a decreasing function of the lattice
size for a given mass. Antiperiodic b.c. have been used. (b.) Same
as in (a), but for a fixed volume
($ V= 1656 $)
and three densities of doping: 20\%, 10\% and 2,4\% .
 $\langle\bar\psi\psi\rangle$  is closest to its regular lattice
value for the smallest density of impurities. (c.) The
value
 $\langle\bar\psi\psi\rangle$  predicted
by sine-Gordon / Thirring (dashed line), including the
fermionic and bosonic contributions,
compared
to the values obtained from the doped lattice (solid line).
The size of the system
is 1600 points and the doping density is 10\%. There are some
uncertainties in the extraction of the masses from the propagators
when $m\to 0$ which may change slightly the dashed line
 but the agreement is qualitatively good.}

\item{Fig. 2.-}(a.) {The fermion effective mass as a function of
the lagrangian mass for a given density (10\%) and three
volumes. For large lagrangian masses the effective mass $M$
is volume
independent, but finite size effects are relevant for small
masses. There is saturation to a finite value that
decreases with increasing volume. (b.) The
fermion effective mass  $M$ for a given volume
($V=1600$) and three densities of impurities. For large masses
the growth of $M$  is linear in $m$ (dotted line), but with a
coefficient
that grows with the doping. The enlarged area shows the departure
from the linear behaviour for small masses (finite size effects)
shown in (a) for a $V=1024$ and different densities.}

\item{Fig. 3.-} (a.) {The scalar zero momentum propagator $P_{11}$
for $V = 800 $, 10\% doping  and $m=0.45$.
The solid line is the doped lattice propagator, the crosses
are the fit assuming that in the scalar channel there is simply
a contribution from a fermion-antifermion pair with mass $2M$
(twice the fermion effective mass) and nothing else.
The wave function renormalization
(which is different from one) is also taken from the fermion
propagator. The fit is perfect. (b.) The
pseudoscalar zero momentum propagator $P_{55}$
for $V = 1600 $, 10\% doping and $m=0.15$.
The dashed line is the fermion-antifermion contribution in a
regular lattice with mass $2M$. The difference between this
and the doped lattice result (solid line) can be accounted for
by a scalar particle (the Goldstone). The crosses are the
combined fit.}

\item{Fig. 4.-} (a.) {$m_{55}$ for a fixed density (10\%) as
a function of the lagrangian mass for three different volumes.
For large masses the finite size effects
are very small. (b.) Same
as before, for a fixed volume
($V=1600$) and three densities.  The coefficient of the
quadratic dependence on $m$ in $m^{2}_{55}$ clearly
grows with the doping (solid line). The fit is extremely good even
for very large masses. For small masses, as seen in (a), there are
departures from this leading $m_{55}\sim m$ behaviour.}

\bye